\documentclass{aa}  

\usepackage{graphicx}
\usepackage{txfonts}
\usepackage{hyperref}
\def\detJ{\mathrm{det}J}

\def\betaein{\beta^*}
\def\Upsilonein{\Upsilon_*^*}

\def\toneobs{\theta_1^{\mathrm{obs}}}
\def\ttwoobs{\theta_2^{\mathrm{obs}}}

\def\moneobs{m_1^{\mathrm{obs}}}
\def\mtwoobs{m_2^{\mathrm{obs}}}

\def\hyperpars{\boldsymbol{\eta}}

\def\Nlens{N_{\mathrm{lens}}}

\def\psilens{\boldsymbol{\psi}_\mathrm{g}}
\def\psilensi{\boldsymbol{\psi}_{\mathrm{g},i}}
\def\psilensnpone{\boldsymbol{\psi}_{\mathrm{g},\Nlens+1}}
\def\psisource{\boldsymbol{\psi}_\mathrm{s}}

\def\psisourcenozs{\boldsymbol{\psi}_\mathrm{s}^{(-z_\mathrm{s})}}
\def\psisourcenozsi{\boldsymbol{\psi}_{\mathrm{s},i}^{(-z_\mathrm{s})}}
\def\psisourcenozsnpone{\boldsymbol{\psi}_{\mathrm{s},\Nlens+1}^{(-z_\mathrm{s})}}

\def\prlens{{\rm P}_\mathrm{g}}
\def\prsource{{\rm P}_\mathrm{s}}

\def\prsl{{\rm P}_\mathrm{{SL}}}

\def\zsource{z_{\mathrm{s}}}
\def\zsourcei{z_{\mathrm{s},i}}
\def\zlens{z_{\mathrm{g}}}

\def\msource{m_{\mathrm{s}}}

\def\pdet{{\rm P}_\mathrm{det}}

\def\data{\mathbf{d}}
\def\datai{\mathbf{d}_i}

\def\nbkg{n_{\mathrm{bkg}}}

\def\mhalo{M_{h}}

\def\rhalo{r_{200}}

\def\reff{R_e}
\def\reffobs{R_e^{(\mathrm{obs})}}
\def\lumobs{L^{(\mathrm{obs})}}
\def\mstar{M_*}

\def\Fref#1{Figure~\ref{#1}\xspace}
\def\Tref#1{Table~\ref{#1}\xspace}
\def\Eref#1{Equation~\ref{#1}\xspace}

\def\pr{{\rm P}}

\defcitealias{S+C21}{Paper~I}
\defcitealias{Son21}{Paper~II}
\defcitealias{Son22}{Paper~III}
\setcitestyle{notesep={}}

\begin{document}

   \title{Statistical strong lensing. IV. Inferences with no individual source redshifts}
   \titlerunning{Statistical strong lensing. IV}
   \authorrunning{Sonnenfeld}


   \author{Alessandro Sonnenfeld\inst{1}
          }

   \institute{Leiden Observatory, Leiden University, Niels Bohrweg 2, 2333 CA Leiden, the Netherlands\\
              \email{sonnenfeld@strw.leidenuniv.nl}
             }

   \date{}

 
  \abstract
    {
Strong lensing mass measurements require the knowledge of the redshift of both the lens and the source galaxy.
Traditionally, spectroscopic redshifts are used for this purpose.
Upcoming surveys, however, will lead to the discovery of $\sim10^5$ strong lenses, and it will be very difficult to obtain spectroscopic redshifts for most of them.
Photometric redshift measurements will also be very challenging due to the blending between lens and source light.
}
   {
The goal of this work is to demonstrate how to carry out an inference of the structural properties of the galaxy population from the analysis of a set of strong lenses with no individual source redshift measurements, and to assess the loss in precision compared to the case in which spectroscopic redshifts are available.
} 
   {
Building on the formalism introduced in Paper III, I developed a method that allows a statistical strong lensing inference to be carried out while marginalising over the source redshifts.
This method, which relies on the knowledge of the properties of the unlensed background source population and of the selection function of the survey, generalises an approach known as photogeometric redshift, originally introduced by the Strong Lensing Legacy Survey collaboration.
I tested the method on simulated data consisting of a subset of 137 strong lenses that is complete above a cut in observational space.
}
   {
The method recovers the properties of the galaxy population with a precision that is comparable to that attainable in the case in which individual source redshifts are known.
}
   {
The photogeometric redshift method is a viable approach for the analysis of large sets of strong lenses provided that the background source population properties and lens selection function are well known.
}

   \maketitle
%

\section{Introduction}\label{sect:intro}

Strong gravitational lensing allows for some of the most precise mass measurements in astrophysics.
The angular separation between strongly lensed images depends on the mass distribution of the lens, as well as on the distances between the observer, the lens, and the background source.
When the redshifts of the lens and source galaxies are known, by modelling the angular structure of the strongly lensed images it is possible to obtain measurements of the total projected mass enclosed within the images with a precision of a few percent, as that is the typical uncertainty associated with the lens modelling step \citep{Bol++08}.

Traditionally, lens and source redshifts are obtained via spectroscopy.
Spectroscopic follow-up observations, however, can be expensive in terms of exposure time. 
In the case of the Strong Lensing Legacy Survey \citep[SL2S; ][]{Ruf++11,Gav++12}, for example, 1-hour-long observations on $8$-metre-class telescopes were needed to obtain the redshift of the background sources.
Moreover, most strongly lensed sources span the redshift range $1-3$ \citep{Son++13b}. A substantial fraction of them do not have any strong emission lines in the optical region of the spectrum, and as a result they go undetected when observed with optical spectroscopy \citep{Ruf++11,Son++13b}.
Near-infrared observations usually guarantee higher chances of success \citep{Son++15,Son++19}, but very few of the existing spectrographs have both the broad wavelength coverage needed to maximise the chances of detecting an emission line and the high spectral resolution required to minimise the contamination from atmospheric emission.

Upcoming surveys such as Euclid\footnote{\url{https://www.euclid-ec.org/}}, the Legacy Survey of Space and Time (LSST\footnote{\url{https://www.lsst.org/}}) and the Chinese Space Station Telescope (CSST) will lead to the discovery of $\sim10^5$ new strong lenses \citep{Col15}.
These lensing data have the potential to enable unique measurements of the mass structure of galaxies.
\citet[][, hereafter Paper I]{S+C21} showed how with a sample of $10^3$ strong lenses it is possible to calibrate stellar mass-to-light ratio measurements of massive galaxies with $0.03$~dex precision.
The analysis of \citetalias{S+C21}, however, was based on the assumptions that the redshifts of all lens galaxies and background sources were known from spectroscopy.
It is unclear whether such spectroscopic data will become available in the near future. 

The standard alternative to spectroscopy is photometric redshift (photo-z) measurements.
The accuracy of photo-zs depends critically on that of the photometric measurements on which they are based.
Most lenses are massive elliptical galaxies, which are bright and with an easily characterised spectral energy distribution. This makes it possible to obtain very precise photometric redshift measurements for them \citep[see for example][]{Vak++19}.
Lensed sources, however, are fainter. Their images have typical $i-$band magnitudes around $24-25$ and are usually found $1-2''$ away from the lens centre \citep{Son++19}. 
Consequently, they are often blended with the light from the lens, which is usually brighter at all but the bluest wavelengths.
This makes it difficult to determine the colours of lensed sources. 
As a result, photometric redshifts have so far found very little use in galaxy-scale strong lensing, with the exception of cases in which high-resolution imaging data in many bands are available \citep{Son++12}.

When it is not possible to obtain individual redshift measurements, the only remaining option is to rely on statistics: if the redshift distribution of the population of background galaxies is known, then it is possible to guess the source redshift of a lens, in a probabilistic sense.
Since lensed sources typically span a broad redshift range, this approach can no longer provide percent mass measurements on individual systems.
Nevertheless, high precision can be gained with the statistical combination of a large number of lenses.

A similar approach is used in weak lensing analyses, where, for the determination of the lensing critical surface mass density, the population distribution of the source redshifts, $\pr(\zsource)$, is used instead of individual photo-z measurements \citep[see for example][]{Hil++17}. 
In the case of strong lensing, however, there is an important complication due to the selection function of the strong lens population.
The probability of a galaxy being strongly lensed by a foreground object varies with its redshift, and so does the probability of it being detected by a strong lens survey.
As a result, the redshift distribution of strongly lensed sources is very different from that of unlensed galaxies: for example, the abundance of sources at a redshift close to that of the lenses is highly suppressed.
Therefore, it is not possible to directly use an estimate of $\pr(\zsource)$ relative to the general background galaxy population as a prior on the source redshift of a strong lens, as that would lead to a biased measurement.

\citet{Ruf++11} tackled this problem in the following way in their study of SL2S lenses.
They assumed that galaxies have an isothermal density profile.
Then, they estimated a prior probability distribution for the lens deflection strength, $\pr(\sigma_{\mathrm{SIE}})$, of the lenses in the SL2S sample by taking the observed stellar velocity dispersion function of galaxies and re-weighting it to take into account the strong lensing cross-section and the luminosity selection of the parent galaxy sample.
The prior probability $\pr(\sigma_{\mathrm{SIE}})$ was then used, together with the observed Einstein radius of each lens, to obtain a posterior probability distribution of the source redshift.
This method of obtaining the redshift of a strongly lensed source is known as photogeometric redshift.

There are two problems with the \citet{Ruf++11} method. The first is that it rests on the assumption of an isothermal density profile for the foreground galaxy population.
The density profile of galaxies, while on average close to isothermal, varies with properties such as stellar mass, half-light radius, and redshift \citep{Son++13b}. Therefore, the isothermal assumption introduces a bias.
While this bias is probably not significant for the \citet{Ruf++11} analysis, it is set to become an important source of systematic errors once precision improves.
More generally, if the purpose of the lensing observations is to determine the structural properties of the galaxy population, then the correct approach is to infer these properties jointly with the redshifts of the background sources, instead of keeping them fixed.
The second problem is that \citet{Ruf++11} did not account for the selection associated with the lens detection phase, which, in general, varies with source redshift.
Both of these problems can be overcome with the formalism introduced in \citet[][, Paper III from here on]{Son22}.

The purpose of this paper is to show how to carry out a population study of a sample of strong lenses with no individual source redshift measurements, and to quantify the loss in precision compared to the case in which spectroscopic redshifts are available.
The method that I propose relies on a good knowledge of the properties of the unlensed background galaxy population, as well as the selection function of the strong lensing survey.
As with \citetalias{Son22}, this is more easily done when the sample is complete above a well-defined cut in observational space.
I tested the method on a simulated population of foreground galaxies and background sources, from which a complete subset of strong lenses was drawn.

The structure of this work is the following.
In Sect. \ref{sect:theory} I introduce the theoretical formalism on which this work is based.
In Sect. \ref{sect:sims} I describe the simulations used to test the analysis method.
In Sect. \ref{sect:model} I describe the model that I used to fit the simulated data.
In Sect. \ref{sect:results} I show the results of the experiment.
I discuss the results and draw conclusions in Sect. \ref{sect:discuss}.

Throughout the paper I assume a flat $\Lambda$ cold dark matter cosmology with $\Omega_M=0.3$ and $H_0=70\,\rm{km}\,\rm{s}^{-1}\,\rm{Mpc}^{-1}$.
Masses and luminosities are in solar units.
The Python code used for the simulation and analysis of the lens sample can be found on GitHub\footnote{\url{https://github.com/astrosonnen/strong_lensing_tools}}.


\section{Photogeometric redshifts}\label{sect:theory}

This section explains (1) how information from strong lensing data, together with prior knowledge of the background source population, can be used to constrain the redshift of a lensed source, and (2) how the properties of a population of galaxies can be inferred by statistically combining data from a subset of lenses, even when their source redshifts are not known.
The following derivation builds on the formalism developed in \citetalias{Son22}.

The population of strong lenses from a survey with selection criterion $S$ can be described with the following probability distribution:
\begin{equation}\label{eq:one}
\prsl(\psilens,\psisource|S) \propto \prlens(\psilens)\prsource(\psisource)\pdet(\psilens,\psisource|S),
\end{equation}
where $\psilens$ and $\psisource$ are parameters describing the foreground galaxy and background source, $\prlens$ and $\prsource$ are the galaxy and source population distribution before lensing, and $\pdet$ is the probability of a lens-source pair being detected as a strong lens by the survey.
The source redshift, $\zsource$, is one of the parameters describing the source. The source parameter array, $\psisource$, can then be written as
\begin{equation}
\psisource = (\psisourcenozs,\zsource),
\end{equation}
where $\psisourcenozs$ includes all of the source parameters (for instance, the magnitude, the position, etc.) except the redshift.

Given a strong lens with observed data $\data$, the marginal posterior probability of the source redshift is
\begin{equation}\label{eq:photogeoz}
\pr(\zsource|\data) \propto \int d\psilens d\psisourcenozs \prsl(\psilens,\psisourcenozs,\zsource|S) \pr(\data|\psilens,\psisourcenozs,\zsource).
\end{equation}
If the strong lens population distribution is known exactly from prior information, then one can use that knowledge to constrain $\pr(\zsource|\data)$ via the above equation.
This is the basic principle of photogeometric redshift.

As the term photogeometric suggests, the constraints on the source redshift consist of a photometric part, which is given by the knowledge, informed by the lens model, of the unlensed surface brightness distribution of the source, and a geometric part, which is directly related to the lens configuration. 
To clarify this concept, it is useful to consider the following example.

We take a strong lens consisting of a foreground galaxy, described by parameters $\psilens$, and a background point source.
The data, $\data$, include the positions and apparent magnitudes of the multiple images of the source.
These data constrain the source redshift via the factor $\pr(\data|\psilens,\psisourcenozs,\zsource)$ in Eq. \ref{eq:photogeoz}, which is the geometric term: modifying the source redshift varies the predicted image positions and magnitudes as a result of the dependence of the lensing deflection angle on the observer-lens-source geometry. 

The photometric constraint on $\zsource$ is instead given by the factor $\prsl(\psilens,\psisourcenozs,\zsource|S)$ in Eq. \ref{eq:photogeoz}.
This can be written as
\begin{equation}
\prsl(\psilens,\psisourcenozs,\zsource|S) = \pr(\psilens,\psisourcenozs|S)\pr(\zsource|\psilens,\psisourcenozs,S),
\end{equation}
where I used the rule of conditional probability.
The first factor in the right-hand side of the above equation is the prior probability of the lens and source parameters (except the source redshift) being $\psilens$ and $\psisourcenozs$ given the fact that the lens is included in the survey with selection criterion $S$.
The second factor, $\pr(\zsource|\psilens,\psisourcenozs,S),$ is the prior probability of the source redshift given the remaining model parameters and the selection criterion.
One of the parameters entering $\psisourcenozs$ is the intrinsic (unlensed) source apparent magnitude, which can be used to further constrain the source redshift given the knowledge of the distribution of background sources in redshift-magnitude space.
For example, if the source is intrinsically very bright, large values of the redshift will be disfavoured.


Source redshift information is rarely needed for its own sake: in most cases, $\zsource$ is required to constrain lens models.
In other words, the factor $\prlens$, which enters Eq. \ref{eq:photogeoz} via the expansion of Eq. \ref{eq:one}, is uncertain, and the source redshift is one of the ingredients needed to constrain it.
\citet{Ruf++11} solved the problem of determining the structural parameters of lenses with no source redshifts as follows.
They used information on $\prlens$ that was independent of their strong lens survey, the SL2S, together with the assumption of an isothermal density profile for the galaxies, to infer the $\zsource$ of sources with no spectroscopic redshifts. 
Then, they combined these estimates of $\zsource$ with the observed Einstein radii and stellar velocity dispersion to infer the evolution of the density profile of the lenses, effectively obtaining an updated estimate of $\prlens$.
Their procedure ensured that no data were used more than once, but it is internally inconsistent because their inferred density profiles of the lenses were different from isothermal. 

The correct approach, when the main goal is to constrain the galaxy population properties, is to infer $\prlens$ while marginalising over the source redshifts.
If we summarise the galaxy distribution with a set of parameters, $\hyperpars$, then we can obtain the posterior probability distribution of $\hyperpars$ given the data as
\begin{align}\label{eq:posterior}
\pr(\hyperpars|\data) \propto \pr(\hyperpars)\prod_i^{\Nlens} \int d\psilensi d\psisourcenozsi & d\zsourcei \pr(\datai|\psilens,\psisourcenozsi,\zsource)\times \nonumber \\
& \times \prsl(\psilensi,\psisourcenozsi,\zsourcei|\hyperpars,S),
\end{align}
where the product is taken over a sample of $\Nlens$ strong lenses.
This is the approach adopted in this paper.

The integrands in the right-hand side of Eq. \ref{eq:posterior} contain the expression for the photogeometric redshift given by Eq. \ref{eq:photogeoz}.
In this case, however, the probability distribution of the lens parameters, $\prsl$, is not fixed but is a function of the galaxy population parameters, $\hyperpars$.
%
If a new lens is observed from a survey with the same characteristics, then its photogeometric redshift can be estimated by marginalising over all possible values of $\hyperpars$ given our knowledge gathered from the analysis of the first $\Nlens$ lenses:
\begin{align}\label{eq:newlens}
\pr(z_{\mathrm{s},\Nlens+1}|\mathbf{d}_{\Nlens+1}) \propto \int d\hyperpars \pr(\hyperpars|\data) \int d\psilensnpone d\psisourcenozsnpone \nonumber \\
\pr(\mathbf{d}_{\Nlens+1}|\psilensnpone,\psisourcenozsnpone) \prsl(\psilensnpone,\psisourcenozsnpone|\hyperpars,S).
\end{align}


\section{Simulations}\label{sect:sims}

In order to demonstrate how the formalism introduced in the previous section can be applied to a practical case, I generated a simulated sample of strong lenses with the following general characteristics.
As in the previous papers of this series, I assumed foreground galaxies to have an axisymmetric mass distribution and background sources to be point-like.
Each galaxy consists of the sum of two concentric components: a dark matter halo and a stellar bulge.
I defined the galaxy population among which strong lenses are searched for by applying a luminosity cut.
The galaxy sample is assumed to be complete above this cut.
Lens-source systems that produce at least two images with observed magnitude brighter than a given threshold are listed as strong lenses detected by the survey.
The survey is assumed to be complete above this limit.
In the next few sections, I provide the details of how this sample was created.

\subsection{Foreground galaxy population}\label{ssec:galmodel}

I generated $10^4$ galaxies, with luminosity, $L$, drawn from the following probability distribution
\begin{equation}\label{eq:lfunc}
\pr(L) \sim \left(\frac{L}{L^*}\right)^{\alpha_L} \exp{\left\{-\frac{L}{L^*}\right\}},
\end{equation} 
with $\log{L^*} = 10.8$, $\alpha_L=-1.0$, and $\log{L} > 10.5$.
Given the luminosity of each galaxy, I obtained its stellar mass by drawing the base-10 logarithm of the stellar mass-to-light ratio, $\Upsilon_*$, from the following distribution:
\begin{equation}\label{eq:upsilon}
\log{\Upsilon_*} \sim \mathcal{N}(0.5,0.1^2),
\end{equation}
where the notation $\mathcal{N}(\mu,\sigma^2)$ indicates a Gaussian distribution with mean $\mu$ and variance $\sigma^2$.
The combination of the luminosity function, Eq. \ref{eq:lfunc}, and the mass-to-light ratio distribution of Eq. \ref{eq:upsilon} produces a stellar mass function that is approximately that of quiescent galaxies at $0.2 < z < 0.5$, as measured by \citet{Muz++13} under the assumption of a \citet{Cha03} stellar initial mass function but shifted towards higher masses by $\sim0.1$~dex.

I described the mass distribution of the stellar component as an axisymmetric de Vaucouleurs profile, with half-light radius drawn from the following stellar mass-size relation:
\begin{equation}
\log{\reff} \sim \mathcal{N}(1.0 + 0.8(\log{\mstar} - 11.5),0.15^2).
\end{equation}
Subsequently, I assigned a dark matter halo mass, $\mhalo$, to each galaxy. 
I defined $\mhalo$ as the mass enclosed within a sphere with average density equal to $200$ times the critical density of the Universe.
I drew the value of $\log{\mhalo}$ from the following Gaussian distribution:
\begin{equation}
\log{\mhalo} \sim \mathcal{N}(13.0 + 1.5(\log{\mstar} - 11.5), 0.2^2).
\end{equation}
The mean of this Gaussian scales with a power $1.5$ of the stellar mass, $\mstar$.
I assumed a spherical Navarro, Frenk, and White \citep[NFW;][]{NFW97} profile for the dark matter distribution, with a fixed ratio between the virial radius and the scale radius, corresponding to a concentration of $5$:
\begin{equation}\label{eq:c200}
\frac{\rhalo}{r_s} = 5,
\end{equation}
where $\rhalo$ is the radius of the sphere enclosing a dark matter mass equal to $\mhalo$.
For simplicity, the foreground galaxies are assumed to be all at the same redshift, $\zlens=0.4$.
The distribution in $\log{L}$, $\log{\mhalo}$, $\log{\Upsilon}$, and $\log{\reff}$ of the sample of foreground galaxies is shown as solid black contours in Fig.\ \ref{fig:sample}.

\subsection{Background source population}\label{ssec:sourcemodel}

I described each background source by means of its $i-$band apparent magnitude, $\msource$, and its redshift, $\zsource$.
I first assumed that their UV absolute magnitude, $M$, is drawn from the following luminosity function:
\begin{equation}
N(M) = 0.4\ln{10}\phi^*\left(10^{-0.4(M - M^*)}\right)^{\alpha_s+1} \exp{\left\{-10^{-0.4(M - M^*)}\right\}},
\end{equation}
with $\phi^*=7.02\times10^{-3}$~Mpc$^{-3}$~mag$^{-1}$, $M^*=-19.68$, and $\alpha_s=-1.32$.
These values correspond to the rest-frame $1500\AA$ luminosity function of $z\approx2$ galaxies, as measured by \citet{Par++16}.

Then, I assumed that all of the sources have the same spectral energy distribution, given by the spectral template `starb1' from the Kinney-Calzetti Spectral Atlas of Galaxies \citep{Cal++94,Kin++96}.
For ease of computation, I only considered galaxies with $\msource < 27$ and $\zlens + 0.1 < \zsource < 4$.
The resulting distribution in $\zsource$ and $\msource$ is shown as solid black lines in Fig. \ref{fig:sample}.
Once integrated over $\zsource$, the background source population has an average number density of $\nbkg=162\,\mathrm{arcmin}^{-2}$.

\subsection{Strong lens definition}\label{ssec:sldef}

Given each foreground galaxy, I randomly placed background sources behind it, assuming a uniform spatial distribution with average projected number density $\nbkg$.
Then, I solved the lens equation to determine their observed image positions.
For simplicity, I assumed the background sources to be massless so that the only source of lensing deflection was that of the foreground galaxy. 
Axisymmetric lenses of the kind considered in this experiment can produce up to three images of the same source.
I focused on the two brightest images, at positions $\theta_1$ and $\theta_2$.
With the label $1$ I indicate the main image, that is, the one outside of the tangential critical curve, which is usually the brightest. Image $2$, instead, is located between the radial and the tangential critical curve, opposite to image $1$ with respect to the lens centre (which is the same notation used in all of the papers of this series).

Given the intrinsic apparent magnitude, $\msource$, of a strongly lensed source and the lensing magnification at the two image positions, I computed the magnitudes of the two lensed images, $m_1$ and $m_2$.
Then, I added a Gaussian observational noise with dispersion $\epsilon_m=0.1$~mag to obtain their corresponding observed values:
\begin{align}
\moneobs & \sim \mathcal{N}(m_1,\epsilon_m^2) \\
\mtwoobs & \sim \mathcal{N}(m_2,\epsilon_m^2).
\end{align}
I defined as a strong lens any galaxy with a strongly lensed source for which the observed magnitude of image $2$ is brighter than $26$.
The lens selection criterion is then
\begin{equation}\label{eq:selcrit}
S = \left\{\mtwoobs < 26 \right\}.
\end{equation}
This condition mimics a selection based on the detection of a counter-image to the main arc, which usually allows the lensing nature of a lens candidate to be determined with high confidence.
I assumed that all such lenses are included in the survey, which is therefore complete above the observational cut given by Eq. \ref{eq:selcrit}.
All of the other galaxies are treated as non-detections.

\begin{figure*}
\includegraphics[width=\textwidth]{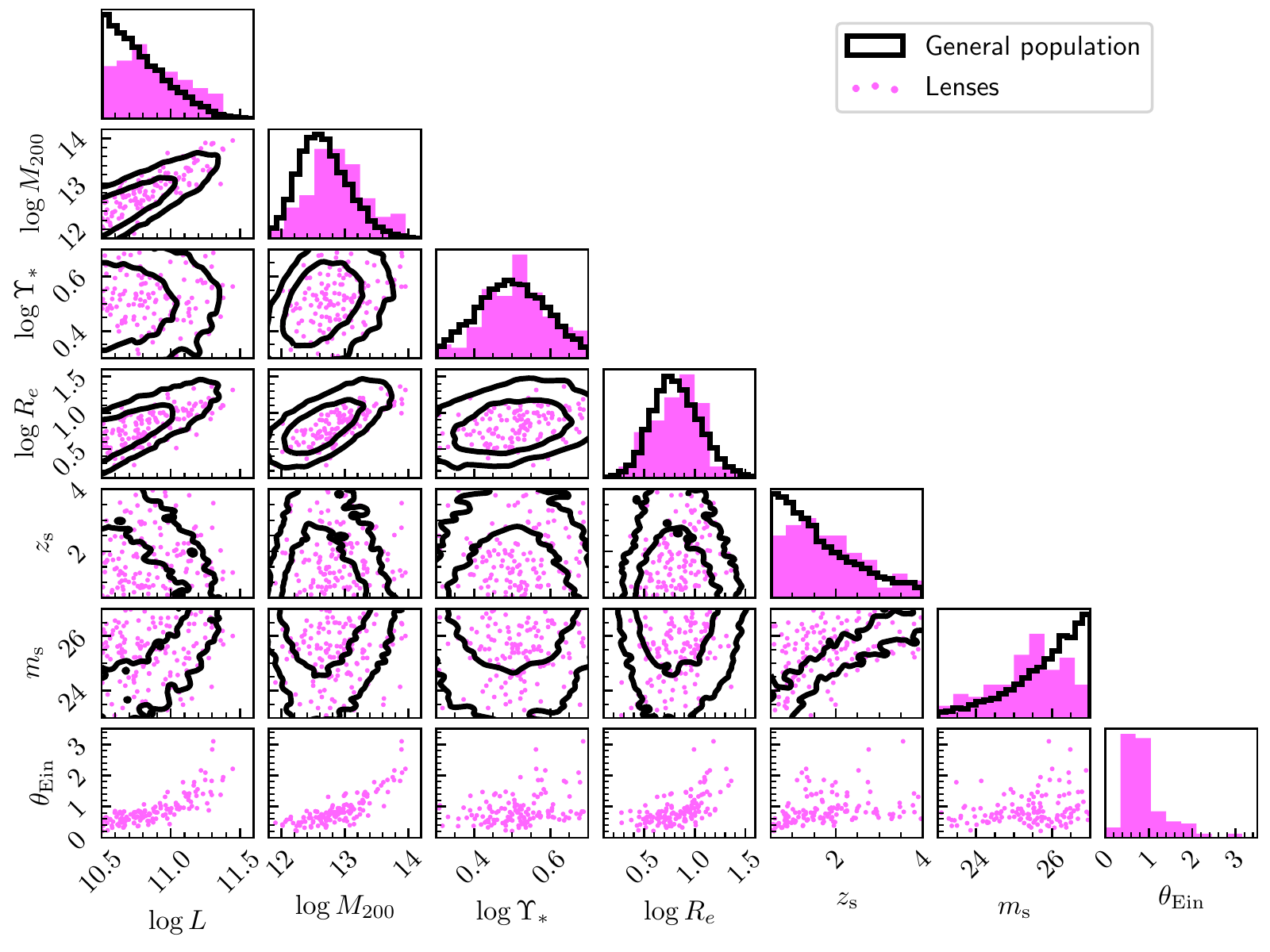}
\caption{
Simulated sample. Solid lines show the distribution in the luminosity, halo mass, stellar mass-to-light ratio, and half-light radius of the general population of foreground galaxies as well as the distribution in redshift and $i-$band magnitude of the general population of background sources. Contour levels correspond to 68\% and 95\% enclosed probability.
Pink points show the values of the above quantities, as well as of the Einstein radius, of the strong lenses.
Histograms are normalised to one.
\label{fig:sample}
}
\end{figure*}

With this definition, the simulation produced $137$ strong lenses.
Two of the lenses have more than one strongly lensed source: in those cases, I only considered one lensed source each, chosen at random.
The distribution in foreground galaxy properties, background source properties, and Einstein radius of the strong lenses is shown in Fig. \ref{fig:sample}.
Compared to the general population, strong lens galaxies tend to be more luminous and, consequently, to have higher halo masses.
Additionally, they are more compact at fixed luminosity (that is, their half-light radius is on average smaller) and have a higher stellar mass-to-light ratio. These last two trends are quantified in \Tref{tab:inference}.

The distribution of strongly lensed sources differs from the general population of background galaxies in two ways.
First, the low-redshift part of the distribution is suppressed. This is because the strong lensing cross-section decreases with source redshift at fixed lens redshift.
Second, the abundance of faint galaxies ($\msource > 26$) is greatly reduced. This is a consequence of the detection limit of Eq. \ref{eq:selcrit}.

\section{Model}\label{sect:model}

I fitted the simulated strong lensing data with a model that describes the joint distribution of the galaxy population, the background source population, and the strong lens detection efficiency of the survey, using the same method introduced in \citetalias{Son22}. Each aspect of the model is described in the following sections.

\subsection{Foreground galaxies}

I described each galaxy with the sum of a stellar component, consisting of a de Vaucouleurs profile with known luminosity and half-light radius and unknown mass-to-light ratio, and a dark matter halo, consisting of an NFW profile.
Moreover, I assumed that the concentration parameter of the NFW profile, Eq. \ref{eq:c200}, is known exactly from prior information.
This is not a realistic assumption, but it is a useful one: it simplifies calculations while still allowing the main goal of this paper to be pursued, which is to demonstrate the accuracy of the inference method itself.

Each galaxy is therefore described by four parameters: the luminosity, $L$, the stellar mass-to-light ratio, $\Upsilon_*$, the half-light radius, $\reff$, and the halo mass, $\mhalo$.
The galaxy parameter set is therefore
\begin{equation}
\psilens \equiv \{L,\Upsilon_*,\reff,\mhalo\}.
\end{equation}
I assumed the following probability distribution for $\psilens$:
\begin{equation}\label{eq:galmodel}
\prlens(\psilens|\hyperpars) = \mathcal{L}(L)\mathcal{U}(\log{\Upsilon_*})\mathcal{R}(\log{\reff})\mathcal{H}(\log{\mhalo}).
\end{equation}
The factor $\mathcal{L}$ is the luminosity distribution of the sample, which I fixed to the true distribution used to generate the mock. 
The factor $\mathcal{U}$ is the distribution in the stellar mass-to-light ratio, which I assumed to be a Gaussian in $\log{\Upsilon_*}$,
\begin{equation}
\mathcal{U}(\log{\Upsilon_*}) \sim \mathcal{N}(\mu_{\Upsilon},\sigma_{\Upsilon}^2),
\end{equation}
with mean $\mu_{\Upsilon}$ and dispersion $\sigma_{\Upsilon}$.

The factor $\mathcal{R}$ is the distribution in half-light radius, which I modelled as the following Gaussian in $\log{\reff}$:
\begin{equation}
\mathcal{R}(\log{\reff}) \sim \mathcal{R}(\mu_{\mathrm{R},0} + \beta_\mathrm{R}(\log{M_*} - \log{M_*^p}),\sigma_\mathrm{R}^2),
\end{equation}
with
\begin{equation}
\log{M_*^p} = 11.0 + \mu_{\Upsilon}.
\end{equation}
In words, the mean $\log{\reff}$ scales with a power $\beta_R$ of the stellar mass, and the pivot point of the stellar mass-size relation, $M_*^p$, is the average stellar mass of a galaxy with luminosity $\log{L} = 11$.
The parameter $\mu_{\mathrm{R},0}$, then, is the average $\log{\reff}$ of a galaxy with luminosity $\log{L}=11$. This quantity can be measured with high precision on the sample of galaxies from which lenses are drawn; therefore, I assume that it is known exactly.

Finally, the factor $\mathcal{H}$ is the distribution in halo mass. Similarly to the $\mathcal{R}$ factor, this is modelled as
\begin{equation}
\mathcal{H}(\log{\mhalo}) \sim \mathcal{N}(\mu_{h,0} + \beta_h(\log{M_*} - \log{M_*^p}), \sigma_h^2).
\end{equation}

The distribution $\prlens(\psilens|\hyperpars)$ is a function of the following set of free parameters:
\begin{equation}\label{eq:hyperpars}
\hyperpars \equiv \left\{\mu_{\Upsilon},\sigma_{\Upsilon},\beta_\mathrm{R},\sigma_\mathrm{R},\mu_{\mathrm{h},0},\beta_\mathrm{h},\sigma_\mathrm{h} \right\}.
\end{equation}
The goal is to infer $\hyperpars$ given the data.
The distribution $\prlens(\psilens|\hyperpars)$ reduces to the true distribution used to generate the mock for
\begin{equation}
\boldsymbol\eta^{(\mathrm{true})} = \left\{0.5,0.1,0.8,0.15,13.0,1.5,0.2\right\}.
\end{equation}

\subsection{Background sources}

Sources are point-like, and therefore they can be fully described with their apparent magnitude, $\msource$, their redshift, $\zsource$, and their position relative to the lens centre, $\beta$:
\begin{equation}
\psisource \equiv \left\{\msource,\zsource,\beta\right\}.
\end{equation}
I assumed that the distribution in $\psisource$ of the background galaxy population is known exactly from prior observations. 
I discuss possible strategies to drop this assumption in Sect. \ref{sect:discuss}.

The probability distribution $\prsource(\psisource)$ is therefore fixed to the one used to generate the mock (shown in black in Fig. \ref{fig:sample}).
This can be written as
\begin{equation}
\prsource(\psisource) = U_{\mathbb{R}^2}(\boldsymbol\beta)\pr(\zsource)\pr(\msource|\zsource).
\end{equation}
In the above equation, $U_{\mathbb{R}^2}$ is a uniform distribution on the sky,  $\pr(\msource|\zsource)$ is
\begin{equation}\label{eq:pmsource}
\pr(\msource|\zsource) \propto \left(10^{-0.4\left(\msource - m_{\mathrm{s}}^*(\zsource)\right)}\right)^{\alpha_s+1} \exp{\left\{-10^{-0.4\left(\msource - \msource^*(\zsource)\right)}\right\}},
\end{equation}
where $\msource^*(\zsource)$ is obtained from $M^*$ via the same spectral template used to generate the mock, 
and $\pr(\zsource)$ is
\begin{equation}\label{eq:pzsource}
\pr(\zsource) \propto \frac{dV_C}{dz} \frac{d \nbkg}{dz},
\end{equation}
which is the product of the derivatives with respect to the redshift of the comoving volume and the surface density of background sources.

\subsection{Lens detection efficiency}

As Sect. \ref{ssec:sldef} explains, the lens sample is complete above the cut of Eq. \ref{eq:selcrit}.
This implies that the detection efficiency of a strong lens, expressed in terms of model quantities, is given by
\begin{equation}
\pdet(\psilens,\psisource|S) = \frac{1}{2}\left[1 - {\mathrm{erf}}{\left(\frac{m_2(\psilens,\psisource) - 26}{\sqrt{2}\epsilon_m}\right)}\right].
\end{equation}

\subsection{Inferring the galaxy population parameters}

I fitted the model described in this section to the data relative to the $137$ strong lenses.
For each lens, I assumed that the following data are available: (1) the positions of the two images of the lensed source, $\toneobs$ and $\ttwoobs$, measured exactly; (2) the half-light radius of the galaxy, $\reffobs$, measured exactly;  (3) the luminosity of the lens galaxy, $\lumobs$, measured exactly; and (4) the magnitudes of the two images, $\moneobs$ and $\mtwoobs$, measured with uncertainty $\epsilon_m$.

In principle, the remaining galaxies in the sample (the non-detections) could be used as additional constraints, following the method introduced in \citetalias{Son22}.
However, \citetalias{Son22} also showed that, when relative magnification information is available (that is, when the magnitudes of the lensed images are used as constraints), the benefit of adding the non-detections is negligible.

Following \citetalias{Son22}, the posterior probability distribution of the galaxy population parameters, given the data of the strong lenses, is given by
\begin{equation}
\pr(\hyperpars|\data) \propto \pr(\hyperpars)\prod_i\pr(\datai|\hyperpars),
\end{equation}
where the product is taken over the strong lenses, and $\datai$ is the data relative to the $i-$th lens.
Each factor $\pr(\datai|\hyperpars)$ can be written as the following integral:
\begin{equation}\label{eq:bigintegral}
\pr(\datai|\hyperpars) = \int d\psilens d\psisource \pr(\datai|\psilens,\psisource) \prsl(\psilens,\psisource|\hyperpars,S),
\end{equation}
where 
\begin{equation}
\prsl(\psilens,\psisource|\hyperpars,S) = A(\hyperpars)\prlens(\psilens|\hyperpars)\prsource(\psisource)\pdet(\psilens,\psisource|S),
\end{equation}
and $A(\hyperpars)$ is a constant that ensures that $\prsl$ is normalised to unity.
The likelihood of observing the data given the lens and source parameters, $\pr(\datai|\psilens,\psisource)$, is
\begin{align}\label{eq:datalike}
\pr(\datai|\psilens,\psisource) = & \delta\left(\reff - \reffobs\right)\delta\left(L - \lumobs\right)\delta\left(\theta_1(\psilens,\beta) - \toneobs\right) \nonumber \\
& \delta\left(\theta_2(\psilens,\beta) - \ttwoobs\right)\pr(\moneobs,\mtwoobs|\psilens,\psisource),
\end{align}
where $\pr(\moneobs,\mtwoobs|\psilens,\psisource)$ is the product of two Gaussians, with means $\moneobs$ and $\mtwoobs$ and dispersion $\epsilon_m$.

\Eref{eq:bigintegral} is a seven-dimensional integral.
The integrals over the galaxy half-light radius and luminosity are trivial, in virtue of the Dirac delta functions of Eq. \ref{eq:datalike}.
Similarly, the fact that the two image positions are known exactly allows the integration to be performed over two lens parameters analytically.
After integrating over the source position $\beta$ and $\log{\Upsilon_*}$, Eq. \ref{eq:datalike} becomes
\begin{align}\label{eq:3dintegral}
\pr(\datai|\hyperpars) = \int & d\zsource d\msource d\log{\mhalo} \pr(\moneobs,\mtwoobs|\mhalo,L\Upsilonein,\reff,\beta^*) \nonumber \\
& \times A(\hyperpars) \pr(L,\reff,\mhalo,\Upsilonein|\hyperpars) \prsource(\zsource,\msource,\betaein) \nonumber \\
& \times \pdet(\psilens,\zsource,\msource,\beta^*|S) \left\lvert\detJ\right\rvert_{\left(\Upsilon_*,\beta\right)=\left(\Upsilonein,\betaein\right)},
\end{align}
where I dropped the subscript $i$ and set $L=\lumobs$ and $\reff=\reffobs$ for ease of notation.
In the above equation, $\Upsilonein$ and $\betaein$ are the values of the stellar mass-to-light ratio and source position needed to reproduce the two image positions,
while the last factor in the integrand is the Jacobian determinant of the following variable change:
\begin{equation}\label{eq:change}
(\log{\Upsilon_*},\beta) \rightarrow (\theta_1,\theta_2).
\end{equation}

I sampled the posterior probability distribution of $\hyperpars$ via a Markov chain Monte Carlo. I assumed a uniform prior on all of the parameters, over the range listed in \Tref{tab:inference}.
At each draw of $\hyperpars$, I computed both the normalisation constant, $A(\hyperpars)$, and the integrals of Eq. \ref{eq:3dintegral} via Monte Carlo integration with importance sampling.

\subsection{The redshift-lens mass degeneracy}\label{ssec:uncertainty}

Before showing the results of the inference over the entire lens population, it is useful to consider the problem of the degeneracy between the source redshift and the lens mass parameters in the case of a single lens.
The lens model used for the fit has five degrees of freedom: the stellar and halo mass of the lens galaxy and the magnitude, redshift, and position of the background source.
If the magnitudes of the two images and the redshift of the source were known exactly (in addition to the image positions), then the model would be fully constrained, and the result of the fit would be equal to the truth.
In the case of unknown source redshift, however, the uncertainty on $\zsource$ propagates to the other lens model parameters.

\begin{figure}
\includegraphics[width=\columnwidth]{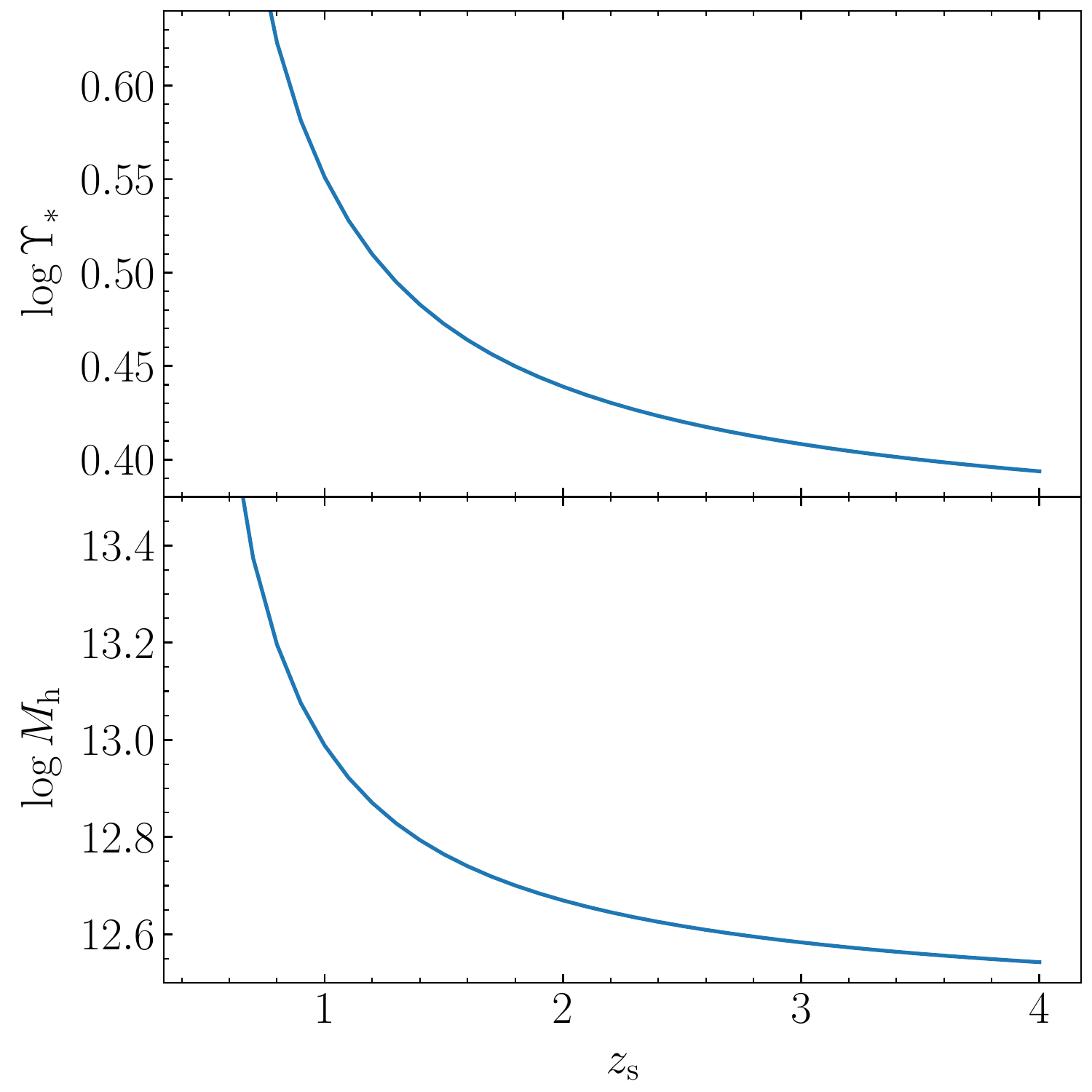}
\caption{
Stellar mass-to-light ratio (upper panel) and halo mass (lower panel) obtained by fitting the lens model introduced in Sect. \ref{ssec:galmodel} to a mock strong lens from the simulated sample as a function of the assumed source redshift. The fit was done assuming perfect measurements of the two image positions and magnitudes.
\label{fig:zserr}
}
\end{figure}

In order to understand the amplitude of this effect, I fitted a lens from the simulated sample, assuming perfect measurements, while assuming a range of values of the source redshift. \Fref{fig:zserr} shows the inferred values of $\Upsilon_*$, $\mhalo$ as a function of $\zsource$.
Both the inferred stellar mass-to-light ratio and the halo mass increase with decreasing source redshift: this is because, as the source gets closer to the lens, the value of the surface critical density for strong lensing becomes larger, and therefore a higher mass is required to produce the deflection angles needed to fit the image positions.
More precisely, when varying the source redshift from $\zsource=1$ to $\zsource=3$, the inferred stellar mass-to-light ratio changes from $\log{\Upsilon_*}=0.55$ to $\log{\Upsilon_*}=0.42$.
While this is a relatively large uncertainty, the combination of a large set of lenses allows this source of noise to be be beaten down statistically, as the next section shows.


\section{Results}\label{sect:results}

\subsection{Galaxy population inference}

\begin{figure*}
\includegraphics[width=\textwidth]{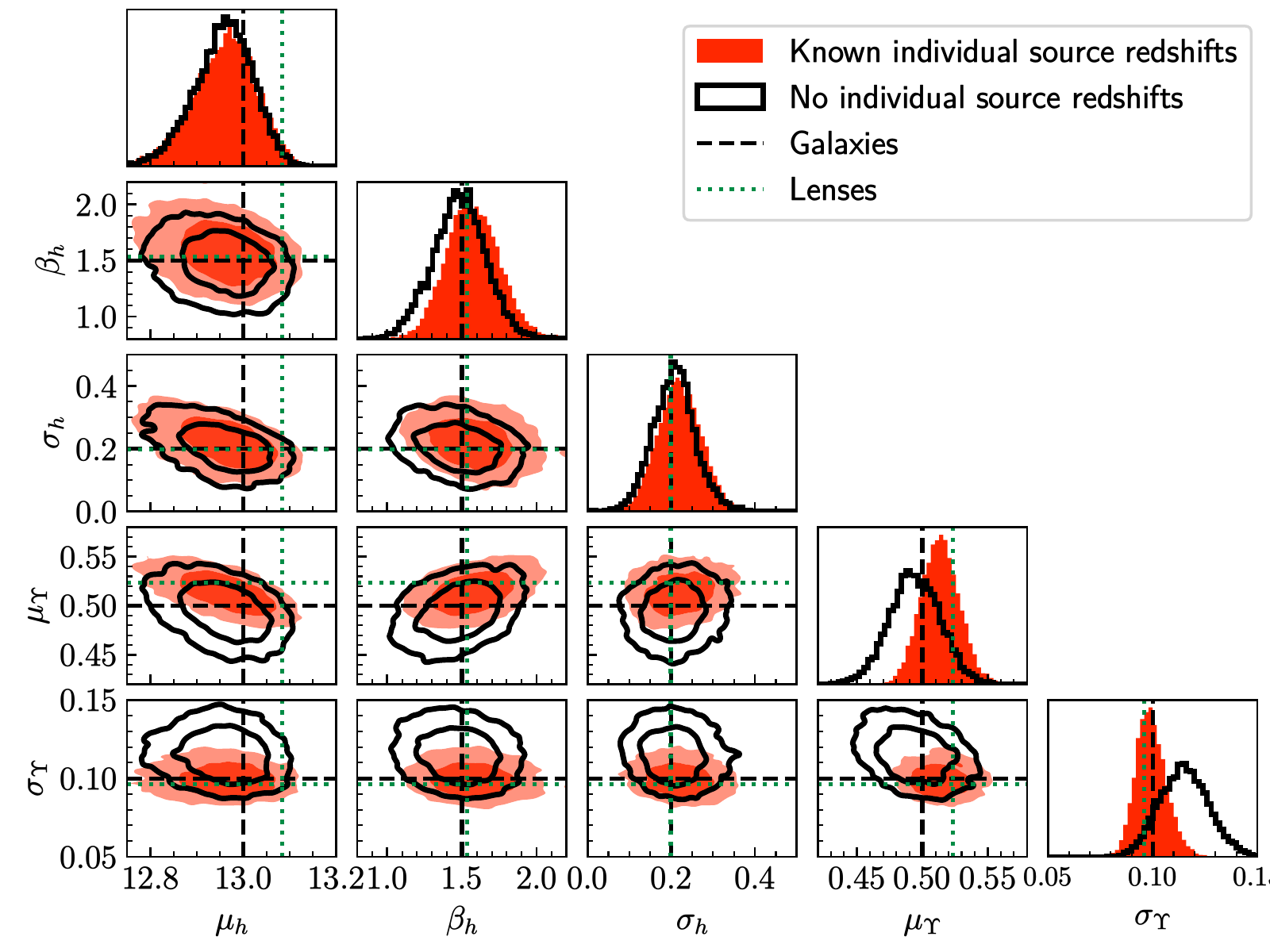}
\caption{
Posterior probability distribution of the parameters describing the distribution in $\mhalo$ and $\Upsilon_*$ of the galaxy population, obtained by fitting the simulated sample of 137 strong lenses with the model of Sect. \ref{sect:model}.
Solid black contours show the inference obtained with no individual source redshift information.
Red filled contours show the inference obtained with known individual source redshifts.
Dashed black lines mark the true values of the parameters.
Dotted green lines indicate the values of the parameters obtained by fitting the model to the subsample of strong lenses.
\label{fig:cp}
}
\end{figure*}

\Fref{fig:cp} shows the posterior probability distribution of the parameters that describe the halo mass and stellar mass-to-light ratio distribution of the galaxy population.
The median and 68\% credible region of the marginal posterior of all of the model parameters (including those describing the size distribution) is reported in \Tref{tab:inference}.
The true values of all of the parameters are accurately recovered.
The inference on the average $\log{\Upsilon_*}$ parameter, in particular, is $\mu_{\Upsilon} = 0.498\pm0.018$.
Moreover, as in \citetalias{Son22}, the model is able to correct for the lensing selection bias: it favours values of the parameters closer to those of the general galaxy population and disfavours values measured on the lenses alone (these values are shown in Fig.\ \ref{fig:cp} and reported in \Tref{tab:inference}), though with low significance.

\begin{table*}
\caption{Inference on the galaxy population model parameters.
}
\label{tab:inference}
\begin{tabular}{ccccccl}
\hline
\hline
Parameter & Truth & Lenses & Prior & No individual & Known individual & Description \\
  & & & & redshifts & redshifts & \\
\hline
$\mu_{\mathrm{h},0}$ & $13.00$ & $13.08$ & $U(12.5,13.5)$ & $12.96\pm0.06$ & $12.96\pm0.07$ & Mean $\log{\mhalo}$ at luminosity $\log{L}=11$ and \\
 & & & & & & average $\Upsilon_*$ \\
$\beta_{\mathrm{h}}$ & $1.50$ & $1.54$ & $U(0.0,3.0)$ & $1.48\pm0.17$ & $1.57\pm0.18$ & Scaling of the mean $\log{\mhalo}$ with $\log{M_*}$ \\
$\sigma_{\mathrm{h}}$ & $0.20$ & $0.20$ & $U(0.0,0.5)$ & $0.21\pm0.05$ & $0.22\pm0.05$ & Scatter in $\log{\mhalo}$ around the mean \\
$\mu_{\Upsilon}$ & $0.50$ & $0.52$ & $U(0.3,0.7)$ & $0.494\pm0.019$ & $0.514\pm0.015$ & Mean $\log{\Upsilon_*}$ \\
$\sigma_{\Upsilon}$ & $0.10$ & $0.10$ & $U(0.0,0.3)$ & $0.115\pm0.012$ & $0.099\pm0.007$ & Scatter in $\log{\Upsilon_*}$ around the mean \\
$\beta_\mathrm{R}$ & $0.80$ & $0.70$ & $U(0.0,2.0)$ & $0.76\pm0.05$ & $0.79\pm0.05$ & Scaling of $\log{\reff}$ with $\log{M_*}$ \\
$\sigma_{\mathrm{R}}$ & $0.15$ & $0.14$ & $U(0.0,0.5)$ & $0.164\pm0.011$ & $0.161\pm0.010$ & Scatter in $\log{\reff}$ around the mean \\
\end{tabular}
\tablefoot{
Column (2): True values of the population parameters. 
Column (3): Values of the lens population parameters, obtained by fitting the model of Eq. \ref{eq:galmodel} to the sample of detected strong lenses.
Column (4): Priors on the parameters.
Column (5): Median and 68\% credible interval of the marginal posterior probability distribution of each parameter.
Column (6): Inference with individual source redshift information.
}
\end{table*}

In order to compare this result with a case in which source redshifts are available, I repeated the analysis by fixing the $\zsource$ of each lens to the true value. The resulting posterior probability distribution is also shown in Fig. \ref{fig:cp} and reported in \Tref{tab:inference}.
The uncertainties obtained when $\zsource$ is known are comparable to the case with no individual $\zsource$ information.
Only the inference on the parameters that describe the stellar mass-to-light ratio is slightly more precise.
This means that the primary factor that determines the uncertainty in this experiment is the limited amount of information that the lensing data provide, and not the lack of source redshift information.
The uncertainties associated with the individual source redshifts are effectively reduced by the statistical combination of the lenses in the sample.

This result depends on the relative importance of the source redshift uncertainty in the error budget.
In turn, this can vary with the quality of the lensing data, the flexibility of the model used to fit the data, the intrinsic scatter in the lens and source population, and the sample size.
Exploring the sensitivity of the results to these various aspects is beyond the purpose of this paper.
We can consider, however, the case in which the two-component galaxy model of Sect. \ref{ssec:galmodel} is replaced with a singular isothermal sphere (SIS).
An SIS is fully described by a single parameter that, in the case in which the source redshift is known, can be constrained exactly with the two image positions of the lensed source.
In other words, uncertainties associated with the individual lens models vanish when assuming an SIS.
As a result, the uncertainty associated with the source redshifts acquires a bigger weight in the error budget, and therefore the difference in precision between the known and unknown individual redshift cases becomes larger.
Nevertheless, massive galaxies are known to have a more complex mass structure than a simple SIS \citep[see e.g.][]{Son++13b}: in all practical cases, when analysing real samples of lenses, models with at least two degrees of freedom are to be preferred.

\subsection{Photogeometric redshifts on new lenses}

\begin{figure}
\includegraphics[width=\columnwidth]{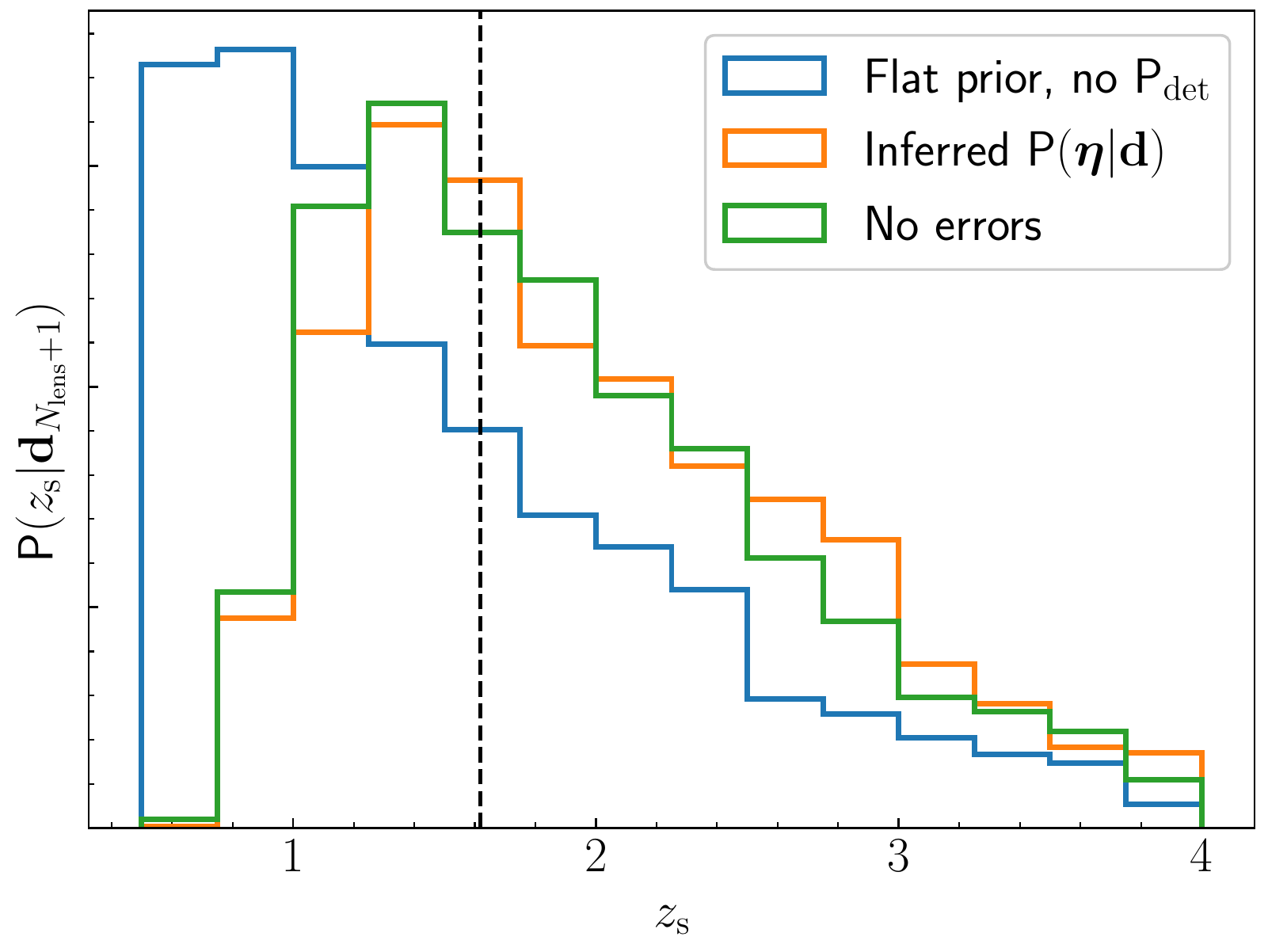}
\caption{
Photogeometric redshift of a strongly lensed source, assuming a lens model from the family of Sect. \ref{ssec:galmodel}.
The blue histogram was obtained by fitting the lensing data assuming a flat prior on $\log{\mhalo}$ and $\log{\Upsilon_*}$ and the source population distribution of Sect. \ref{ssec:sourcemodel}, with no correction for the strong lensing selection.
The orange histogram was obtained assuming the full model, including the selection function correction, and using the values of the galaxy population parameters, $\hyperpars$, inferred in the experiment as a prior.
The green histogram was obtained neglecting observational errors and fixing the values of $\hyperpars$ to the truth.
The dashed vertical line marks the true value of the source redshift.
\label{fig:newlens}
}
\end{figure}

Once the properties of the population of galaxies have been determined, it is possible to use this knowledge to predict the source redshifts of individual strong lenses.
We can consider a new lens, discovered by a survey with identical properties to the one on which the mock sample was based.
The marginal posterior probability of the source redshift of this lens is given by Eq. \ref{eq:newlens}. \Fref{fig:newlens} shows this posterior probability for an example lens, fitted with the same model of Sect. \ref{sect:model} and using the posterior probability of the galaxy population parameters, $\pr(\hyperpars|\data)$, obtained earlier.

In order to illustrate the way in which the knowledge of the galaxy population informs this quantity, Fig. \ref{fig:newlens} also shows the marginal posterior probability on $\zsource$ obtained (1) assuming flat priors on the lens parameters $\log{\mhalo}$ and $\log{\Upsilon_*}$ and (2) neglecting the lens detection efficiency term, $\pdet$.
The two distributions are very different: the information on the galaxy population properties and on the selection function correctly disfavours solutions with source redshifts very close to the lens. On the contrary, when this information is not taken into account, the posterior probability in the source redshift stays similar to the prior distribution of the background source population, which is highest at the lowest redshifts.

The width of the $\pr(\zsource|\mathbf{d}_{\Nlens+1})$ distribution is the result of (1) uncertainties associated with the lens model, (2) uncertainties on the galaxy population distribution parameters, $\hyperpars$, and (3) the width of the $\prsl(\psilens,\psisource|\hyperpars)$ distribution.
In order to isolate these different sources of uncertainty, I repeated the analysis assuming perfect lensing measurements (that is, assuming no errors on the magnitudes of the two images) and fixing $\hyperpars$ to the true value.
The resulting posterior probability distribution, also shown in Fig. \ref{fig:newlens}, is very similar to the one obtained without these additional assumptions.
This means that the main sources of uncertainty on the photogeometric redshift are the width of the galaxy and source population distributions.

%


\section{Discussion and summary}\label{sect:discuss}

The analysis described in the previous sections relied on the assumption of a perfect knowledge of the distribution of the background source population, $\prsource(\psisource)$. 
As discussed in \citetalias{Son22}, it can be challenging to acquire this information in practice, especially if the selection criteria of a survey involve more aspects than just the magnitudes and positions of the lensed images.
For example, if the detection probability is also a function of surface brightness, then the model $\psisource$ must also include parameters such as galaxy size.

In principle, $\prsource(\psisource)$ can be determined by studying the population of unlensed galaxies.
However, because of lensing magnification, the appearance of strongly lensed sources, and therefore their detectability, can vary with changes in $\psisource$ that occur below the resolution limit of the observations with which the unlensed galaxy population is studied.
A possible solution, then, is to determine $\prsource(\psisource)$ along with the galaxy population, adding flexibility to the model.
This will introduce an additional source of uncertainty.
In that case, the redshift dependence of $\prsource(\psisource)$ could be constrained by obtaining spectroscopic redshifts for a small subset of the background sources in the strong lens sample.
I leave the exploration of the case in which the source population distribution term is no longer fixed for future study.

This paper revisited the concept of photogeometric redshift, addressing the theoretical shortcomings present in the original formulation of \citet{Ruf++11}.
The experiment carried out demonstrates that it is possible to use strong lenses with no individual source redshift information to make accurate inferences of the properties of the galaxy population.
When dealing with a large sample of lenses ($\Nlens > 100$), the statistical noise from the uncertainty on the individual source redshifts is greatly reduced. In the case in which the properties of the background (unlensed) source population are known exactly, its contribution to the error budget on the galaxy population parameters is only secondary.

The key elements that are necessary for a successful application of this method are (1) the knowledge of the lens selection function and (2) the knowledge of the distribution in surface brightness profile-redshift space of the background galaxy population. The problem of determining the former can be greatly simplified by identifying a subset of the strong lens population that is complete above an observational cut, as shown in \citetalias{Son22}. Measuring the latter is potentially made difficult by finite resolution effects, the importance of which depends on the specific method used to discover lenses. Nevertheless, the distribution of the source properties can be inferred along with the properties of the foreground galaxy population by extending the model.
In summary, the method introduced in this paper provides a viable solution to the problem of analysing large samples of lenses when spectroscopic follow-up is not possible.



\begin{acknowledgements}
I thank Phil Marshall and Raphael Gavazzi for useful suggestions.
\end{acknowledgements}

\bibliographystyle{aa}
\bibliography{references}

\end{document}